# Fe-H melting curve below 3 GPa: Implications for hydrogen in the lunar core

J. Takeshita[1], K. Hirose[1,2]*, S. Fu[3], F. Sakai[1], K. Hikosaka[1,4]

1. Department of Earth and Planetary Science, The University of Tokyo, Tokyo 113-0033, Japan
2. Earth-Life Science Institute, Institute of Science Tokyo, Tokyo 150-8550, Japan
3. School of Earth Sciences, Zhejiang University, Hangzhou 310058, China
4. Department of Earth and Planetary Sciences, Institute of Science Tokyo, Tokyo 150-8551, Japan

* Corresponding author (email: kei@eps.s.u-tokyo.ac.jp)

**Abstract**

It has been assumed that hydrogen is negligibly incorporated into core-forming metals below ~3 GPa, and therefore the presence of hydrogen in iron cores of small terrestrial bodies including the moon has not been considered. Here we performed high-pressure melting experiments on the Fe-H system under $H_2$-saturated conditions, combined with synchrotron X-ray diffraction (XRD) measurements. Results demonstrate substantial depression of the Fe-H melting curve compared to that for Fe at 1.0–3.3 GPa, indicating that hydrogen is incorporated into liquid iron even at low pressures less than 1 GPa and the solubility is enhanced with increasing pressure. Based on the density of liquid Fe-H derived from diffuse scattering signal in XRD data, we found that the solubility of hydrogen in liquid iron is about 0.9 wt% at 3.6 GPa and likely enhanced to 1.2 wt% at 5 GPa corresponding to lunar core conditions. The 1.2 wt% H causes 9% density reduction, which might fully explain the observed density deficit of the lunar core with respect to iron, depending on the density estimate from seismological data.

**Introduction**

Geophysical constraints on the lunar interior indicate that the density of its core is a few to ~40% lower than that of pure iron (*e.g.,* Garcia *et al.*, 2019; Viswanathan *et al.*, 2019; Kuskov *et al.*, 2021; Zhao *et al.*, 2023). Such density deficit requires the presence of substantial amounts of light elements in the metallic core, and sulfur and carbon have often been regarded as possible light alloying elements (*e.g.,* Jing *et al.*, 2014; Steenstra *et al.*, 2017). In contrast, hydrogen has received comparatively less attention since stoichiometric FeH is formed from Fe and H$_2$ only above 3.5 GPa at room temperature (Badding *et al.*, 1991). Indeed, hydrogen was assumed to be least soluble into iron below ~3 GPa in recent core-formation modelling (Tagawa *et al.*, 2021; Tsutsumi *et al.*, 2025).

Quite few high-pressure and -temperature (*P-T*) experiments were previously reported on the Fe-H system below 3 GPa. All neutron diffraction measurements on Fe-H alloys have been carried out at pressures of ~3 GPa, and at temperatures far below the onset of melting (*e.g.,* Machida *et al.*, 2019; Iizuka-Oku *et al.*, 2017; Ikuta *et al.*, 2019). Yagi and Hishinuma (1995) performed XRD measurements on Fe-H and determined melting temperatures at pressures down to 2.2 GPa, demonstrating a large reduction from the melting curve of Fe. Other experimental studies reported the Fe(-Ni)-H melting temperatures only above 2.7 GPa (Suzuki *et al.*, 1984; Okuchi, 1998; Fukai *et al.*, 2003; Sakamaki *et al.*, 2009; Shibazaki *et al.*, 2014; Mita *et al.*, 2025).

In this study, we focus on Fe-H melting experiments below ~3 GPa to explore the enhancement of hydrogen solubility in iron with increasing pressure. The results show the clear melting temperature depression even at 1.0 GPa and higher hydrogen concentrations in liquid and solid iron at higher pressures, suggesting that hydrogen could be an important light element in metallic iron cores of small terrestrial bodies. We discuss the possible hydrogen abundance in the lunar core based on seismological constraints on its density.

**Results**

We conducted melting experiments on the Fe + H$_2$ sample at four different *P-T* conditions (laser heating was performed at two separate portions of a sample in each of two independent runs) (Table S-1). In run #1, after compression under 300 K to 1.1 GPa, we first heated the sample at spot #1-1 and observed diffuse scattering signal indicative of melting, along with the XRD peaks from face-centred cubic (fcc) FeH$_{0.16}$, when temperature was increased from 1490 K at 1.3 GPa to 1530 K at 1.5 GPa (Fig. 1). After quenching to 300 K, we decompressed the sample to 0.6 GPa and then heated a fresh sample portion (spot #1-2) being exposed to X-ray and laser beams. XRD data exhibited the diffuse scattering signals from liquid coexisting with fcc FeH$_{0.08}$ upon increasing temperature from 1590 K to 1640 K at 1.0 GPa. Subsequently, we decreased the sample temperature to 1550 K and lost such diffuse signals (Fig. 1). The XRD data were also collected at higher temperature of 1710 K, providing a diffuse scattering signal sufficient to determine the liquid density.

Similarly in run #2, we initially compressed the sample to 1.8 GPa and observed diffuse scattering upon increasing temperature from 1360 K at 2.1 GPa to 1440 K at 2.2

GPa, coexisting with fcc FeH$_{0.22}$ (spot #2-1). The diffuse scattering signals disappeared when decreasing temperature from 1440 K to 1310 K. Subsequently the sample was further compressed to 2.7 GPa, and a fresh sample area was heated. The diffuse scattering signals appeared when increasing temperature from 1230 K at 3.0 GPa to 1380 K at 3.3 GPa (spot #2-2), coexisting with fcc FeH$_{0.29}$. We also collected an XRD pattern from this spot at 1670 K at 3.6 GPa for the analysis of liquid density.

These results locate the melting (solidus) curve of Fe-H in a pressure range from 1.0 to 3.3 GPa, showing that its melting temperature decreases monotonically from ambient pressure to ~3 GPa (Fig. 2), rather than a sudden reduction above ~2 GPa (Yagi and Hishinuma, 1995; Fukai *et al.*, 2003). Earlier studies attributed such melting temperature drop to a change in the subsolidus phase from body-centred cubic (bcc) to fcc at ~2 GPa (Fukai *et al.*, 2003; Shibazaki *et al.*, 2014), but the present experiments demonstrate that liquid Fe-H coexists with fcc and the melting temperature is depressed already at 1.0 GPa.

We obtained hydrogen concentrations $x$ in both liquid and coexisting fcc FeH$_x$ from XRD data collected at temperatures above the melting (solidus) curve (Fig. 3, Table S-1). The estimate of hydrogen abundance in the solids was based on the observed volume expansion of FeH$_x$ with respect to Fe (see Experimental methods in Supplementary Information). For the liquids, we first estimated the density of liquid FeH$_x$ from diffuse scattering signals (Kuwayama *et al.*, 2020; Fu *et al.*, 2025) and then obtained its hydrogen abundance considering the volume expansion by hydrogen; hydrogen concentration in liquid FeH$_x$ increased from $x$ = 0.14 at 1.0 GPa to 0.48 at 3.6 GPa. The increase in the hydrogen abundance in liquid with increasing pressure is supported by the shift of the first-peak position in structure factor $S(Q)$ (Fig. S-1). Since experiments were performed under H$_2$-saturated conditions, the resulting values correspond to the hydrogen solubility limit in liquid iron at each pressure.

**Discussion**

*Melting phase diagram of Fe-H at low pressures*    The present experiments demonstrate that the melting temperature in the Fe-H system decreases with increasing pressure from <1 GPa to ~3 GPa (Fig. 2). The magnitude of depression from the Fe melting curve is enhanced almost linearly with increasing pressure; ~240 K at 1 GPa, ~450 K at 2 GPa, and ~580 K at 3 GPa. It is explained by a linear increase in the solubility of hydrogen primarily into liquid iron with increasing pressure to ~3 GPa as evidenced by the XRD measurements (Fig. 3).

The pressure evolution of the Fe-H melting phase diagram is illustrated in Figure S-2, in which we consider that the hydrogen solubility into liquid iron (excess hydrogen is present as liquid $H_2$) increases linearly with increasing pressure to ~3 GPa, and accordingly the depression of eutectic temperature (above which metallic liquid is formed) from the Fe melting curve is also linearly enhanced with pressure. Earlier experiments reported that the Fe-H melting temperature conversely rises above ~3.5 GPa (Yagi and Hishinuma, 1995; Okuchi, 1998) (Fig. 2). It could be a consequence of a rapid increase in the hydrogen solubility into fcc Fe (Fig. S-2), which is inferred from the formation of stoichiometric FeH above that pressure at 300 K (Badding *et al.*, 1991).

***Possible hydrogen abundance in the lunar core***   The present experiments demonstrate the depression of melting temperature of iron by ~240 K even at 1.0 GPa in the presence of excess $H_2$, suggesting that liquid iron can incorporate hydrogen not at ambient pressure but certainly under lower pressures than previously thought. While recent planetary accretion and core formation modelling assumed that hydrogen is not soluble into core-forming metals below 3 GPa (Tagawa *et al.*, 2021; Tsutsumi *et al.*, 2025), it is possible that metallic cores of small rocky bodies such as our moon also include some hydrogen. Indeed, the moon should have been covered with the protolunar disk gas, and therefore a lunar magma ocean (LMO) likely included a certain amount of water, most of which could have been later incorporated as hydrogen into core-forming iron metals.

The depth of the LMO has been estimated to be 600 km, which explains the crustal thickness when all plagioclase crystallizing from the LMO formed the crust (Charlier *et al.*, 2018). It is possible that the lunar core formation (in other words, core metal segregation from silicate) took place at its bottom under 2.8 GPa. On the other hand, Rai and van Westrenen (2014) found that core metal may have reached chemical equilibrium with silicate at 4.5±0.5 GPa, which most likely matches the pressure at the core-mantle boundary. We consider the representative lunar core pressure to be 5 GPa. The present-day core temperature may be about 1700 K (Garcia *et al.,* 2019; Kuskov *et al.*, 2021), which is well above the solidus temperature of the Fe-H system at 5 GPa (Fig. 2).

The solubility of hydrogen in liquid iron at 2.8 GPa or 5 GPa may limit its concentration in the lunar core. Considering that the reduction in Fe-H melting temperature with increasing pressure to ~3 GPa is primarily caused by the pressure-induced enhancement of hydrogen solubility into liquid iron (see above), the hydrogen content in liquid iron at the melting (solidus) temperature will correspond to the solubility limit. Our results (Fig. 3) suggest that the hydrogen solubility limit in liquid iron may be $x$ = 0.37 and 0.65 in $FeH_x$ (0.7 and 1.2 wt% H) at 2.8 and 5 GPa, respectively.

We can also calculate the possible hydrogen content in the lunar core based on its metal-silicate partitioning under core formation conditions. Water concentration in the bulk silicate moon (BSM) has been estimated from the analyses of lunar volcanic glasses. Hauri *et al.* (2015) found that the BSM is similar in highly volatile element composition to the depleted MORB source mantle of the Earth, giving 133–292 ppm $H_2O$ in the BSM. The magma ocean models by Elkins-Tanton and Grove (2011) demonstrated that the LMO should have contained 100 to >1000 ppm water when the later melting source region for such volcanic glasses includes 10 to 200 ppm water. Also, Lin *et al.* (2020) argued that the lunar crustal thickness (the amount of plagioclase) is controlled by water abundance in the LMO, which may be ~300 ppm $H_2O$ if the LMO was 700 km deep. Under plausible conditions of the lunar core formation of 2.8 GPa, 2163 K, and oxygen fugacity relative to the iron-wüstite buffer $\Delta IW = –2$ (Rai and van Westrenen, 2014), the partition coefficient of hydrogen $D_H^{metal/silicate}$ is $1.11 \times 10^2$ (molar basis) according to Tsutsumi *et al.* (2025) (Fig. S-3). Even with 300 ppm $H_2O$ (33 ppm H) in the LMO (Lin *et al.*, 2020), the metal-silicate partitioning predicts 0.3–0.4 wt% H in the lunar core. If the LMO included 1000 ppm $H_2O$ (111 ppm H) at the time of the core formation, core hydrogen concentration could be as high as 1.1–1.2 wt% H (see Calculation of metal-silicate partitioning of hydrogen in Supplementary Information).

***Is hydrogen a major light element in the lunar core?*** The density of the lunar core has been modelled to be 4200–5200 kg/m$^3$ by Garcia *et al.* (2019) and 5560–6070 kg/m$^3$ by Viswanathan *et al.* (2019) assuming homogeneous liquid core. The more recent work by Kuskov *et al.* (2021) proposed a higher density range of 6200–7000 kg/m$^3$ for the liquid core, considering the outer liquid and inner solid cores (Fig. 4).

We calculated the density of liquid Fe-H by considering the volume expansion due to the incorporation of hydrogen atom into iron (Fig. 4). The maximum 0.7–1.2 wt% H in the lunar core based on the hydrogen solubility limit reconciles within uncertainty the lunar liquid core density deficit estimated by Kuskov *et al.* (2021) with hydrogen alone, although it is not enough for the lower densities proposed earlier by Garcia *et al.* (2019) and Viswanathan *et al.* (2019). Furthermore, if the metal-silicate chemical equilibrium was attained at the bottom of the LMO, our estimate of 0.3–0.4 to 1.1–1.2 wt% H in the core can explain the observed density within its large uncertainty (Kuskov *et al.*, 2021).

These results suggest that hydrogen could be, at least, a major light element in the lunar core. Sulfur and carbon have previously been proposed to be important lunar core light elements, but we note that their solubilities into solid iron coexisting liquid alloy are limited only to <0.5 wt% S and 1.5 wt% C at 5 GPa (Li *et al.*, 2001; Fei and Brosh, 2014).

This indicates that if the lunar solid inner core also exhibits a certain density deficit with respect to iron, it is attributed not to sulfur nor carbon but to hydrogen.

**Acknowledgements**

We thank H. Kadobayashi and N. Hirao for their assistance in the synchrotron experiments and high-pressure gas loading at SPring-8 (proposals no. 2023B0306, 2025A1133 and 2025B1212). This work was supported by the JSPS grant 21H04968 to K. Hirose.

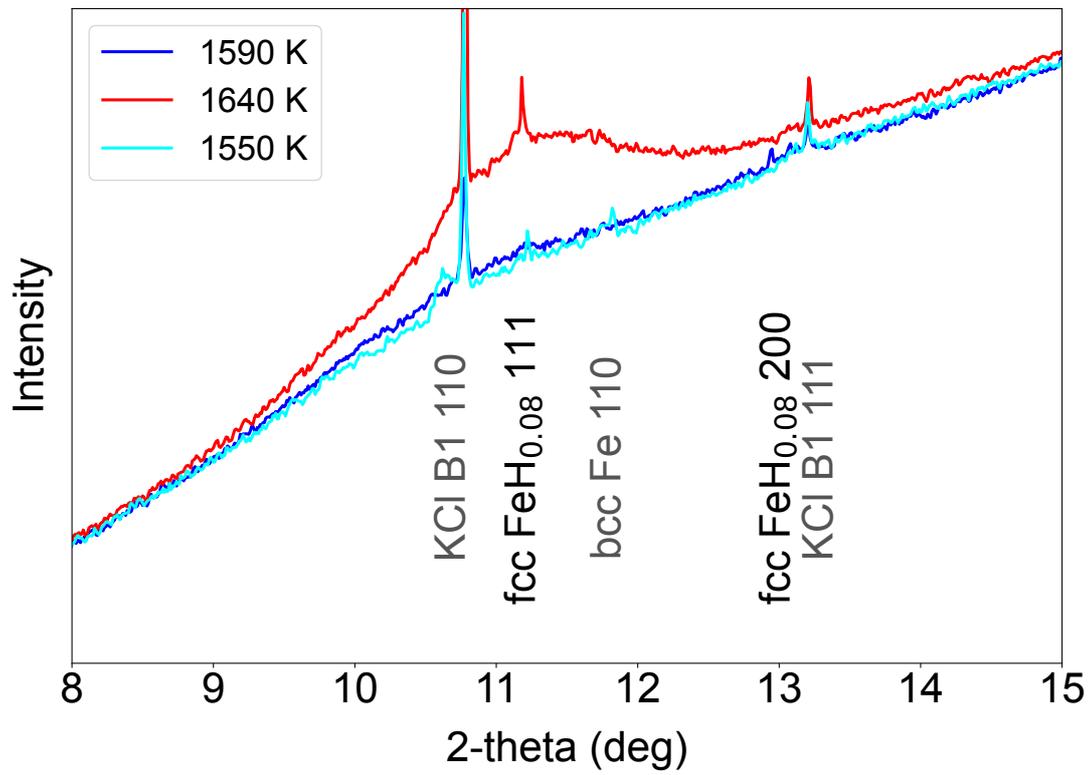

**Figure 1** XRD patterns obtained from run #1 (spot #1-2) at 1.0 GPa. Diffuse scattering indicative of melting appeared when sample temperature was increased from 1590 K (blue) to 1640 K (red) and then disappeared upon cooling to 1550 K (cyan).

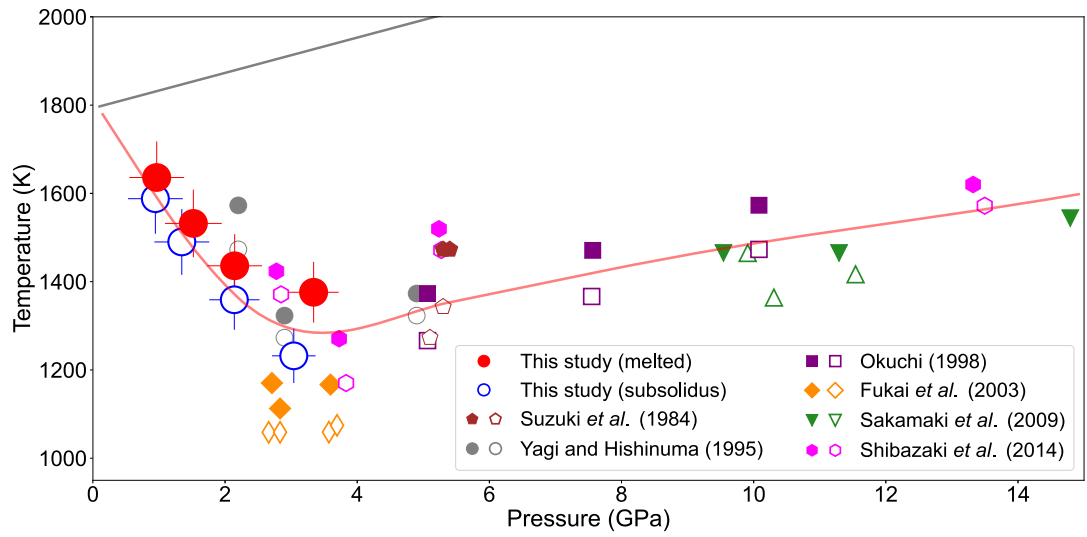

**Figure 2**  Melting curve of FeH$_x$ ($0 < x < 1$) (red line). The grey line represents the melting curve of pure Fe.

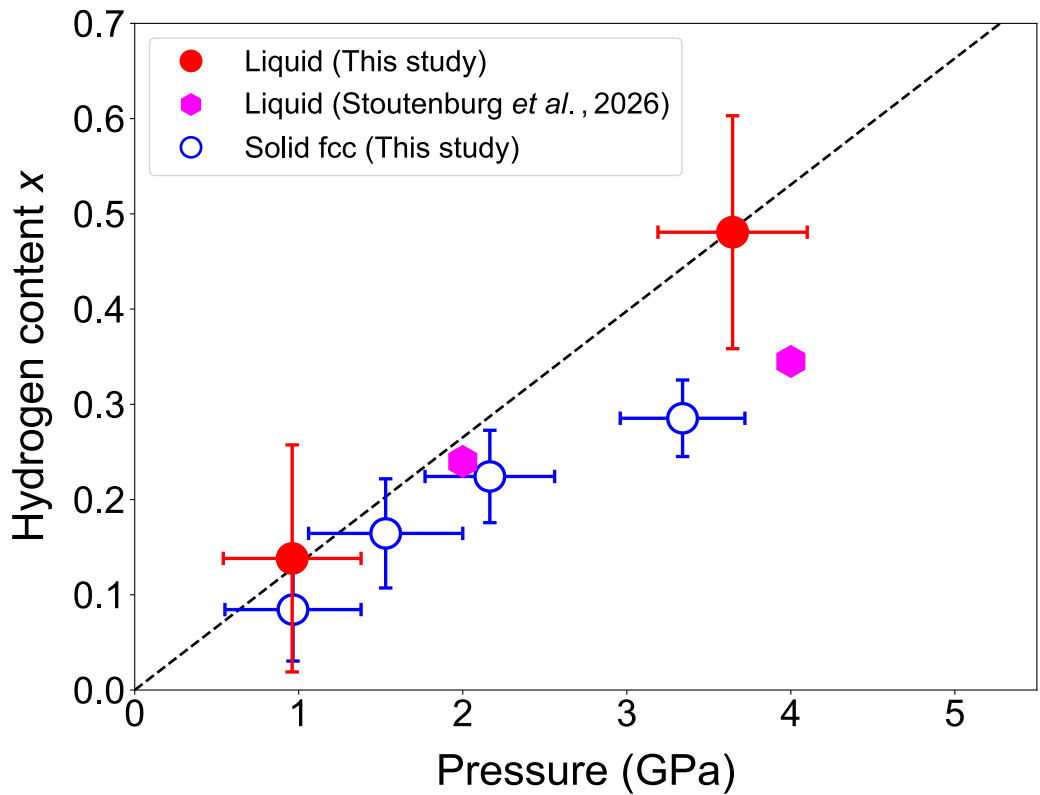

**Figure 3** The hydrogen contents in coexisting solid fcc and liquid phases in this study. The theoretical predictions of those in liquid by Stoutenburg *et al.* (2026) at 1500 K are also shown.

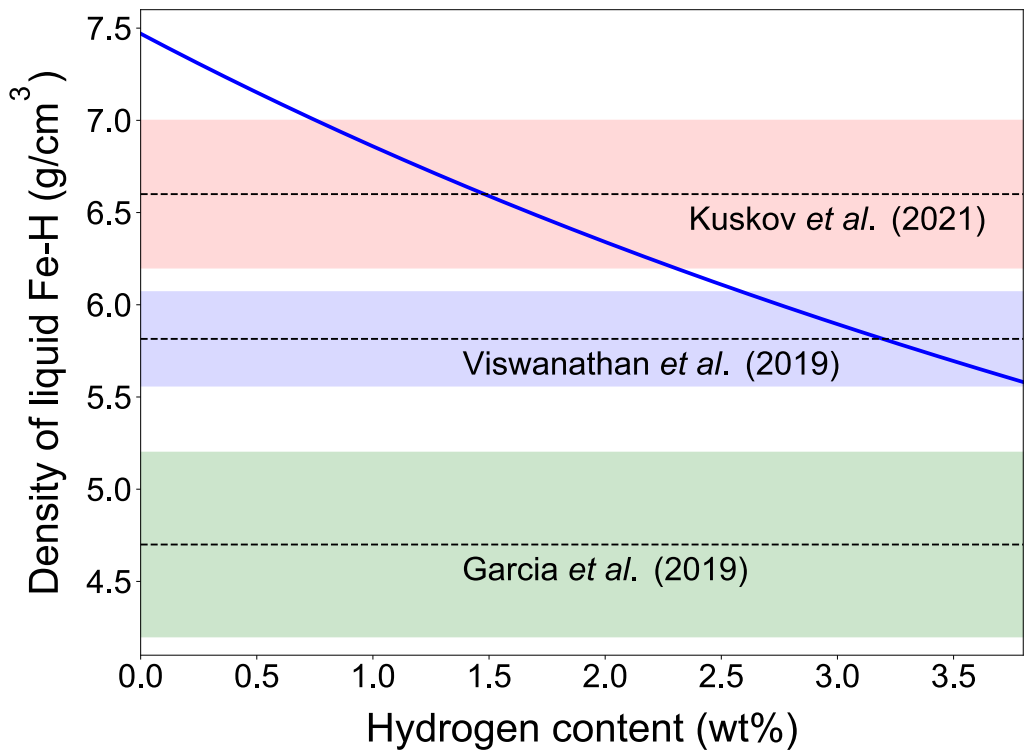

**Figure 4** Density of liquid FeH$_x$ as a function of hydrogen concentration at 5 GPa and 1700 K. The density of liquid pure Fe is from Kuwayama *et al.* (2020), and density reduction is calculated with $\Delta V_H$ = 2.22 Å$^3$ (volume expansion by one H atom) (see Experimental methods in Supplementary Information). The solubility limit of hydrogen into liquid iron is 0.7 wt% at the base of the LMO and 1.2 wt% at at the core-mantle boundary. Metal-silicate partitioning suggests 0.3–0.4 to 1.1–1.2 wt% H in the core, depending on hydrogen concentration in the LMO. These amounts of hydrogen account for the lunar core density estimated by Kuskov *et al.* (2021), while other light elements such as sulfur and carbon are required for earlier estimates by Garcia *et al.* (2019) and Viswanathan *et al.* (2019). See text for details.

# Fe-H melting curve below 3 GPa: Implications for hydrogen in the lunar core

J. Takeshita, K. Hirose, S. Fu, F. Sakai, K. Hikosaka

## Supplementary Information

The Supplementary Information includes:

- Experimental methods
- Calculation of metal-silicate partitioning of hydrogen
- Table S-1
- Figures S-1 to S-4
- Supplementary Information References

## Experimental methods

High-pressure and -temperature (*P-T*) experiments were conducted in a laser-heated diamond-anvil cell (DAC) using diamond anvils with flat 800 µm culet (we can precisely control experimental pressure by using anvils with such large culet size). The anvil surface was coated with Ti by sputtering in order to prevent hydrogen penetration into diamonds (Ohta *et al.*, 2015). We loaded a pure Fe foil (Toho Zinc, >99.999% purity) with a thickness of approximately 20 µm into a hole at the centre of a pre-indented Re gasket, leaving space for hydrogen. Pressure markers of KCl and a ruby ball were placed beneath the Fe foil. Subsequently the DAC was placed in an oven maintained at 120 °C for 1 hour in order to remove moisture from the KCl. Supercritical hydrogen fluid was introduced into the sample chamber using a high-pressure gas apparatus (PRETECH Co., Ltd) at SPring-8. The high-pressure vessel and gas line were evacuated for 5 minutes and subsequently purged four times with $H_2$ gas prior to loading into the DAC. The presence of $H_2$ molecules was confirmed by Raman spectra. The sample was then compressed to a target pressure, with monitoring pressure by the ruby ball.

We obtained X-ray diffraction (XRD) data at the beamline BL10XU of SPring-8. The incident X-ray beam was monochromatized to ~30 keV and focused to 6 µm in diameter (full-width at half maximum). XRD patterns were collected using a flat panel X-ray detector (Varex Imaging, XRD1611CP3). The obtained 2D XRD images were converted into 1D patterns using IP Analyzer software (Seto *et al.*, 2010), and phase identification and unit-cell volume derivation were performed. We heated the Fe + $H_2$ sample from both sides using a couple of 100 W single-mode Yb fibre lasers (SPI photonics). The laser spot size was approximately 50 µm in diameter. Using visible fluorescence that appeared upon irradiating X-ray onto the KCl medium, the laser beams from both sides were adjusted to be coaxial with the X-ray beam. The laser power output was increased stepwise, and the XRD spectrum and temperature were collected/measured each time. The laser power output was maintained for several tens of seconds at each step. Temperature was determined using a spectro-radiometric method (Hirao *et al.*, 2020). The sample temperature was taken as the average within the 6 µm area probed by the XRD beam. The temperature uncertainty may be ±5%. Pressure under high temperature was determined from the lattice volume of the B1 or B2 phase of KCl and their thermal equations of state (EoSs, Walker *et al.*, 2002; Tateno *et al.*, 2019). We adopted its effective temperature as $T_{\text{KCl\_effective}} = (3 \times T_{\text{KCl\_max}} + 300)/4$ following Campbell *et al.* (2009), in which $T_{\text{KCl\_max}}$, the maximum temperature in KCl, could be as low as 300 K but as high as measured sample temperature (or the melting temperature of KCl when it is lower). Errors in pressure thus primarily derived from the large uncertainty in $T_{\text{KCl\_max}}$.

It is known that hydrogen escapes from iron when it turns to the bcc phase (*e.g.,* Iizuka-Oku *et al.*, 2017; Tagawa *et al.*, 2021). It is necessary to determine hydrogen concentration $x$ in fcc FeH$_x$ at high *P-T* because it transforms into the bcc phase upon quenching temperature. The hydrogen content is calculated as:

$$x = \frac{V_{\text{FeH}_x} - V_{\text{Fe}}}{\Delta V_{\text{H}}} \tag{S-1}$$

where *V* represents the volume per chemical formula (half of the unit-cell volume of fcc). $V_{\text{Fe}}$ for the fcc phase was derived from the EoS of pure fcc-Fe reported by Tsujino *et al.* (2013). $\Delta V_{\text{H}}$ (= 2.22 Å$^3$), the volume expansion upon incorporation of one hydrogen atom into the Fe lattice, was adopted from the value for fcc FeH$_x$ at 4–12 GPa and about 1000 K reported by Ikuta *et al.* (2019).

The amount of hydrogen in liquid iron could not be obtained in the manner described above, because the liquids crystallized into bcc Fe upon quenching temperature, which does not incorporate hydrogen into its lattice. Alternatively, we obtained the density of liquid FeH$_x$ and then determined $x$ considering $\Delta V_{\text{H}}$. First, liquid density was calculated from the diffuse scattering signal in XRD data following the method recently developed by Kuwayama *et al.* (2020) (Fig. S-4). High-pressure XRD measurements in a DAC always have a limitation on the range of momentum transfer $Q$ (up to $Q_{\text{max}}$) because of a geometrical constraint, which leads to artificial oscillations in distribution function $F(r)$ and pair distribution function $g(r)$. Indeed, no atoms can exist at distances shorter than the nearest-neighbour atomic distance ($r < r_{\text{min}}$), meaning that the local density must be zero. We therefore complemented the data to satisfy such physical requirement, $g(r) = 0$ at $r < r_{\text{min}}$ (in other words, for $F(r)$ to exhibit a straight line with a slope of $-4\pi\rho r$, where $\rho$ is the average atomic number density). The derived density typically includes ±~2% uncertainty. More technical details are found in Kuwayama *et al.* (2020) and Fu *et al.* (2025), who derived the densities of liquids Fe and FeH$_x$, respectively. The hydrogen content in liquid FeH$_x$ is then estimated from the density difference from liquid pure Fe and $\Delta V_{\text{H}} = 2.22$ Å$^3$ from Ikuta *et al.* (2019). The density of liquid Fe is from the EoS by Kuwayama *et al.* (2020), which employed the data at ambient pressure as well. According to the EoSs of liquid Fe by Kuwayama *et al.* (2020) and liquid Fe-H by Fu *et al.* (2025), the densities of Fe and FeH$_{0.3}$ are 7.47 and 7.13 g/cm$^3$, respectively, at 5 GPa and 1700 K, indicating $\Delta V_{\text{H}} = 2.17$ Å$^3$, which matches the value by Ikuta *et al.* (2019).

## Calculation of metal-silicate partitioning of hydrogen

The hydrogen content in the lunar core was calculated using $D_\text{H}^\text{metal/silicate}$ (molar basis) reported by Tsutsumi *et al.* (2025). We adopted the lunar primitive upper mantle composition from Longhi (2003) for the LMO composition and 0 to 7 wt% S in the core following Jing *et al.* (2014). Note that Tsutsumi *et al.* (2025)'s $D_\text{H}^\text{metal/silicate}$ values were obtained by considering that hydrogen does not have a colligative property in iron solvent since small H atoms are incorporated into liquid Fe interstitially and therefore the activity of element *i* in metal is approximated as;

$$x'_i = \frac{N_i}{\sum_{k \neq \text{H}} N_k} \tag{S-2}$$

## Supplementary Tables

**Table S-1** Experimental results on melting and hydrogen content $x$ in FeH$_x$.

| Run | | #1 | | #2 | |
|---|---|---|---|---|---|
| Spot | | #1-1 | #1-2 | #2-1 | #2-2 |
| Subsolidus | $P$ [GPa] | 1.3(4) | 1.0(4) | 2.1(4) | 3.0(3) |
| | $T$ [K] | 1490(70) | 1590(80) | 1360(70) | 1230(60) |
| Right above solidus | $P$ [GPa] | 1.5(5) | 1.0(4) | 2.2(4) | 3.3(4) |
| | $T$ [K] | 1530(80) | 1640(80) | 1440(70) | 1380(70) |
| | $x$ (fcc) | 0.16(5) | 0.08(6) | 0.22(5) | 0.29(4) |
| Liquid density measurements | $P$ [GPa] | – | 1.0(4) | – | 3.6(4) |
| | $T$ [K] | – | 1710(90) | – | 1670(80) |
| | $\rho$ (liquid) [g/cm$^3$] | – | 7.03(12) | – | 6.87(12) |
| | $x$ (liquid) | – | 0.14(12) | – | 0.48(12) |

# Supplementary Figures

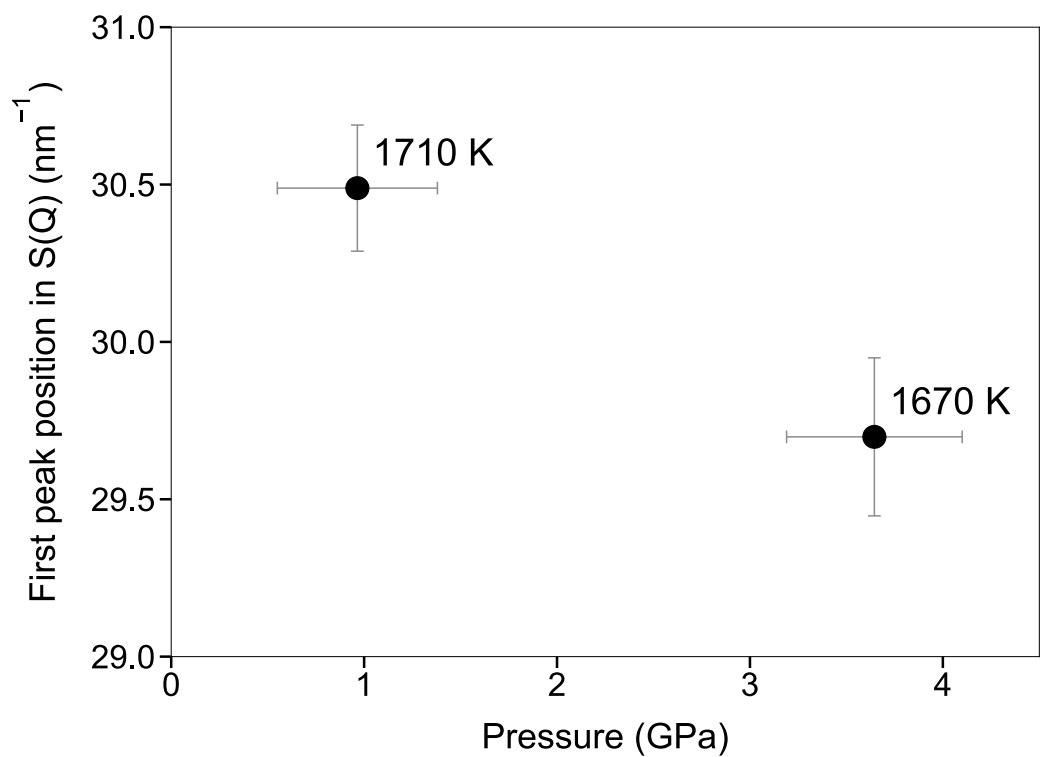

**Figure S-1** The first peak position in the structure factor $S(Q)$. The observed shift indicates a larger volume and thus higher hydrogen concentration with increasing pressure from 1.0 to 3.6 GPa.

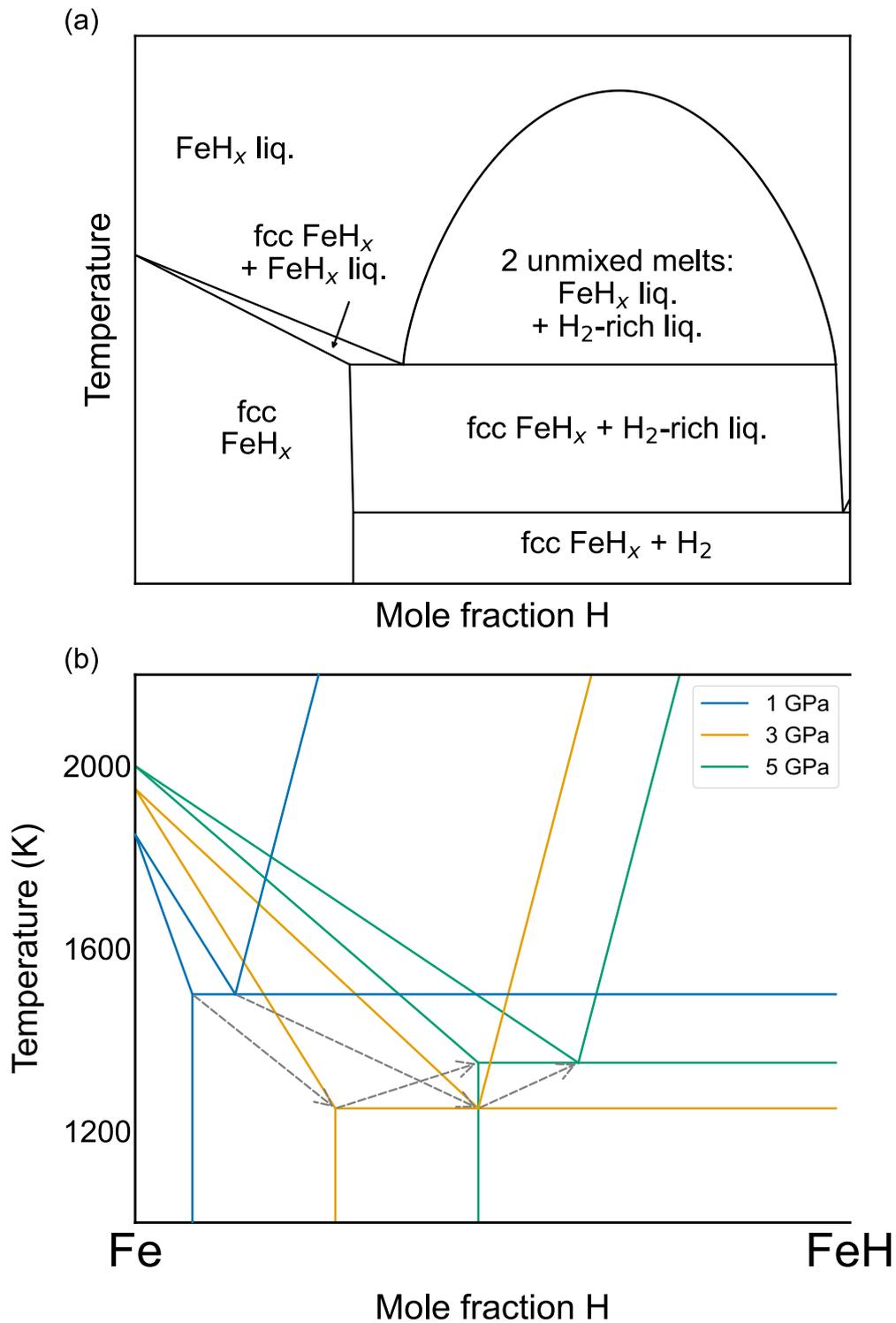

**Figure S-2** (a) Schematic phase diagram between Fe and $H_2$ under pressure (less than ~10 GPa) based on Tagawa *et al.* (2022) and Stoutenburg *et al.* (2026). (b) The Fe-rich portion of the Fe-H phase diagram, showing the enhanced solubilities of hydrogen in solid and liquid iron and the changes in solidus temperature with increasing pressure from 1 to 5 GPa.

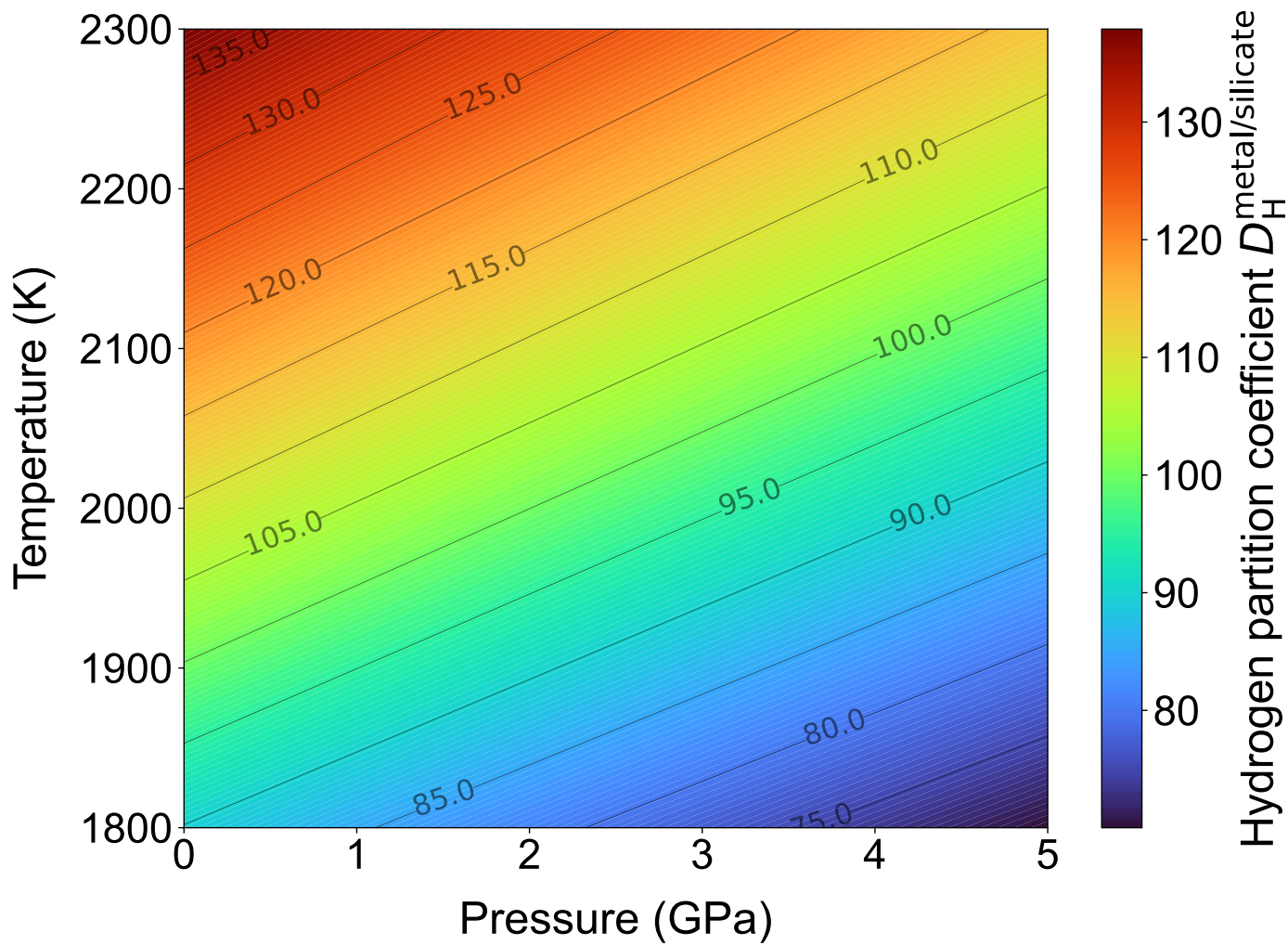

**Figure S-3** Metal-silicate partition coefficient of hydrogen ($D_H^{metal/silicate}$, molar basis) as functions of pressure and temperature from Tsutsumi *et al.* (2025), considering ΔIW = –2 (Rai and van Westrenen, 2014) and carbon-free metal composition.

**Figure S-4** Determination of liquid density from diffuse scattering using the analytical method developed by Kuwayama *et al.* (2020). **(a)** The structure factor $S(Q)$, **(b)** reduced interference function $f(Q)$, **(c)** distribution function $F(r)$, and **(d)** radial distribution function $g(r)$ ($Q$ is momentum transfer). The black lines represent raw data, and the blue broken lines show complements added for the physical requirement that no atoms can exist at distances shorter than the

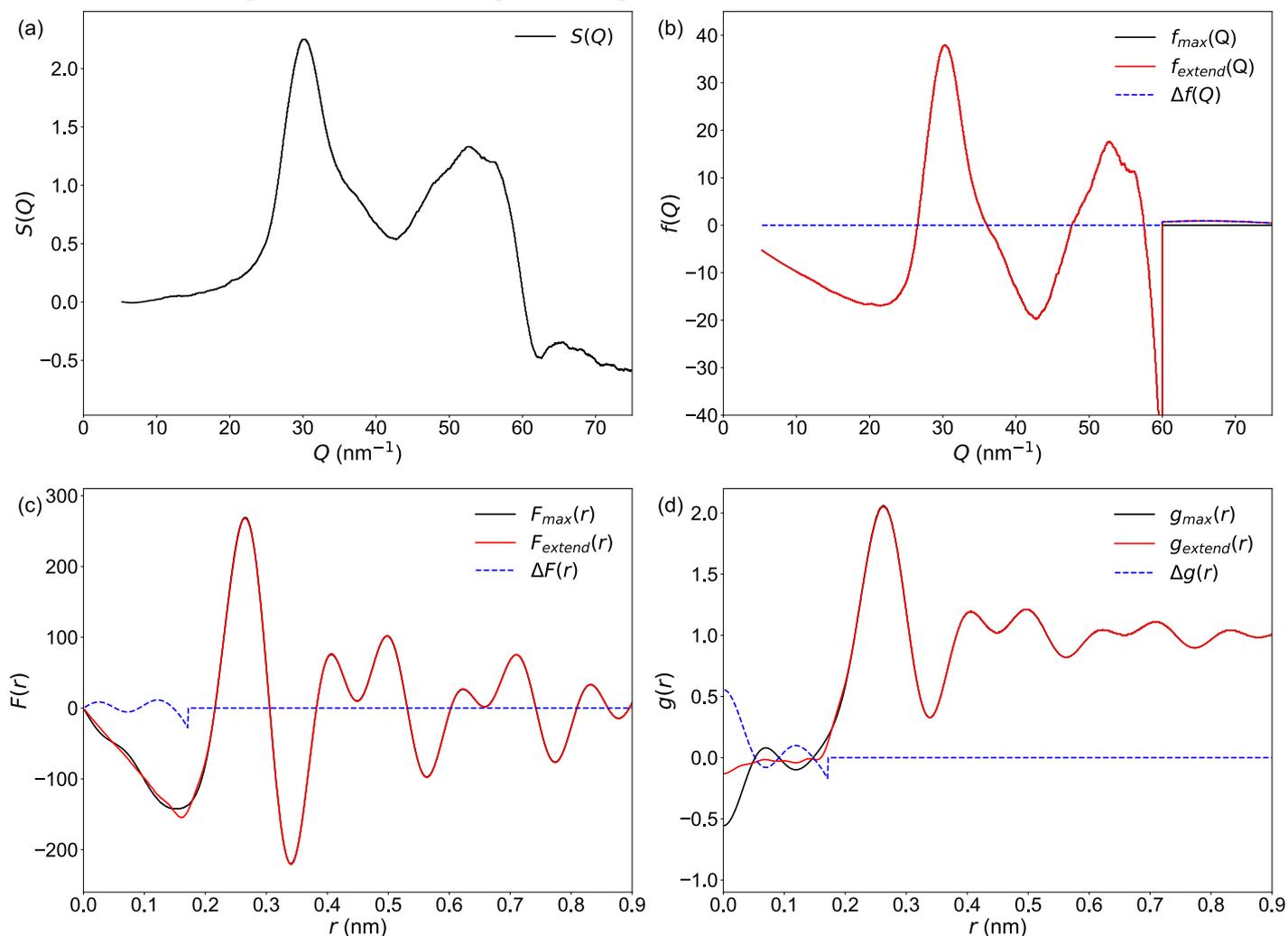

nearest-neighbour atomic distance [$g(r) = 0$ at $r < r_{min}$]. With such data complements (red lines), $F(r)$ exhibits a straight line with a slope of $-4\pi\rho r$ ($\rho$ is average atomic number density). See the Supplemental text for details.

# Supplementary Information References